\documentclass{ws-ijmpaMY}
\usepackage{hyperref}

\begin{document}
\markboth{A. A. Grib \& Yu. V. Pavlov}{Fifty years of the calculation of the density
of particles created in the early Universe}

%
%

\title{Fifty years of the calculation of the density of particles created in\\
the early Friedmann Universe}

\author{A. A. Grib}

\address{A.~Friedmann Laboratory for Theoretical Physics, Saint Petersburg, Russia;\\
Theoretical Physics and Astronomy Department of the Herzen  University,\\
Moika 48, Saint Petersburg 191186, Russia\\
andrei\_grib@mail.ru}

\author{Yu. V. Pavlov}

\address{Institute of Problems in Mechanical Engineering, Russian Academy of Sciences,\\
Bol'shoy pr. 61, St. Petersburg 199178, Russia;\\
N.I.\,Lobachevsky Institute of Mathematics and Mechanics,
    Kazan Federal University,\\
    18 Kremlyovskaya St., Kazan 420008, Russia\\
yuri.pavlov@mail.ru}

\maketitle

\begin{history}
\received{Day Month Year}
\revised{Day Month Year}
\end{history}

\begin{abstract}
    The problem of particle creation in cosmology concerning whether
the results for the number of particles are infinite or finite is discussed
for scalar and spinor particles in Friedmann expanding Universe.
It is shown that the results are always finite if one puts in case of scalar
particles creation a special term into the Lagrangian.
Numerical estimates of the effect of particle creation are made.
The role of creation of superheavy particles in cosmology is discussed.

\keywords{Particle creation; early Universe; curved space-time.}
\end{abstract}

\ccode{PACS numbers: 04.62.+v, 95.35.+d, 98.80.Cq}


\section{Introduction}
\label{secI}

    In Ref.~\refcite{GribNuclPhys69} published by one of the present authors
(A.A.G.) together with S.G.~Mamayev 50 years ago finite results for the space
density of conformal massive scalar particles created in the Friedman expanding
space were obtained.
    These results were later confirmed in
the original papers\cite{GribNuclPhys71,GribMamayevMostep76,MMStarobinsky} and
in the monograph.\cite{GMM80}
    Finite results for this effect were also obtained for spin one half
massive particles.\cite{Mamayev76}
    Particle creation in cosmology was also considered in the papers of
Ya.B.~Zel'dovich and A.A.~Starobinskii
(see Refs.~\refcite{Zeldovich70,ZeldovichStarobinsky71})
and the monograph.\cite{Zeldovich75}

    However differently from our papers in the paper of
L.~Parker\cite{Parker69} and Ref.~\refcite{BD} infinite results
for the creation of minimally coupled to gravitation scalar particles
were obtained and due to this fact the authors claimed that there are some
irresistible difficulties in the whole problem of particle creation
in cosmology.
    Finite results for the minimally coupled scalar particle creation were
obtained by Yu.V. Pavlov.\cite{Pavlov00}

    To solve the controversy we give in this paper all necessary formulas
explicitly showing when and why the results are finite or infinite.
    Surely infinite results cannot have physical sense.

    Let in the neighborhood of some space-time point $M$ the invariant
of the curvature tensor has the value of the order
    \begin{equation}
R_{ijkl} R^{ijkl} \sim \rho^{-4},
\end{equation}
    where $\rho$ is the characteristic curvature radius.
    One can introduce a coordinate system that will be locally Galilean up to
distances of the order of~$\rho$.
    One can construct in this system a complete set of one-particle functions,
which for the frequencies $\omega_\alpha \gg \rho^{-1}$ will be,
with the exponential accuracy, positive- and negative-frequency relative to
time coordinate.
    However, for the frequencies
\begin{equation}
\omega_\alpha \le  \rho^{-1}
\end{equation}
the difference between positive- and negative-frequency functions, generally
speaking, disappears, that corresponds to the unity order uncertainty for the
number of particles in the mode~$\alpha$.

    In analogy with electrodynamics particle creation can be understood as
breaking of vacuum loops by the external gravitation force of
the expanding Universe.
    Gravity equally act on particles and antiparticles and breaking of
vacuum loops is explained by the work of tidal forces.
    To define the breaking forces consider the geodesic deviation equation
\begin{equation}
\frac{d^2 n^i}{d s^2 } = R^i_{\, jkl} u^j n^k u^l .
\end{equation}
     Let's $ u^0 =1$, $u^\alpha=0$, \ $n^0 =0$, $|n^\alpha | \sim l_C = m^{-1}$,
i.e. as characteristic distance between particles we take Compton length $l_C$.
    For particle creation one must have that work of tidal forces at
the distance of the order of $l_C$ would exceed $2m$
\begin{equation}
| R^\alpha_{\, 0 \beta 0}| \ge l_C^{-2} \ge m^2 .
\end{equation}
    So the curvature of space–time must be of the order of the inverse
Compton length.

    Massless particles due to the property that Friedmann space is
conformally static and the equations for truly massless particles must be
conformal invariant are not created.
    But for large enough curvature one can neglect mass and this means that
only for the curvature corresponding to the Compton length there is an era
of particle creation.
    Roughly speaking two particles are created on the Compton length and
surely the particle density in space must be finite.

    It is also simple to see that created particle number in any finite volume
of the Friedman Universe is proportional to the number of causally disconnected
parts of the universe at the time of creation.
    This argument partly was proposed to the author (A.A.G.) by Ya.B.~Zel'dovich
in private meeting in St. Petersburg (at that time Leningrad) in 1970.

\section{Wave Equations in Curved Space-Time}
\label{secWE}

     Free scalar field with mass $m$ in curved space-time satisfies equation
\begin{equation}
( {\nabla}_i {\nabla}^i + \xi R + m^2 )\, \varphi(x)=0 ,
\label{mEq}
\end{equation}
    where $\nabla_i $ is covariant derivative, $R$ is the scalar curvature.
    If $\xi=0$ this is a field with minimal coupling.
    For $\xi=\xi_c =1/6 $ it is the field with conformal coupling.

    By the conformal transformation of the space-time metric $g_{ik}(x)$
and the field
    \begin{equation}
g_{ik} \to {\tilde g }_{ik} = \exp [-2\sigma(x)] \, g_{ik} ,
\label{cgik}
\end{equation}
    \begin{equation}
\varphi(x) \to \breve{\varphi}(x) = \exp [ \sigma(x) ] \, \varphi(x),
\label{phitilde}
\end{equation}
    where $\sigma(x)$ is an arbitrary smooth function of coordinates,
the equation~(\ref{mEq}) for $\xi=\xi_c$ is transformed into
    \begin{equation}
\left( {\tilde \nabla}_i {\tilde \nabla}^i + \xi_c {\tilde R} +
m^2 e^{2 \sigma} \right)  \breve{\varphi}(x) =0 ,
\label{CPSP}
\end{equation}
    where ${\tilde \nabla}_i$ and ${\tilde R}$ are calculated
in metric ${\tilde g}_{ik}$.
The Eq.(\ref{mEq}) is conformally invariant, if  $m=0, \ \xi=\xi_c$.

    The equation for the free field with spin one half in curved space-time is
\begin{equation}
\left( i\, \gamma^k(x) \overrightarrow{\mathstrut \nabla}_{\!k} - m \right)
\psi(x) =0 \,,
\label{Diraceq}
\end{equation}
    where $ \gamma^k(x) = h_{(a)}^{\, k}(x)\, \gamma^a,$
$\gamma^a $ are Dirac matrices, $ h_{(a)}^{\, k}(x) $ are tetrad vectors,
      \begin{equation}
\overrightarrow{\mathstrut \nabla}_{\!k} \psi =
\left( \partial_k +(1/4) C_{abc} h^{(c)}_{\, k} \gamma^b \gamma^a \right) \psi,
\ \ \ \
C_{abc} = \left( \nabla_{\! i}\, h_{(a)}^{\, k} \right) h_{(b) k}.
\label{gamx}
\end{equation}

    The Dirac equation~(\ref{Diraceq}) is conformally invariant for $m=0$,  if
together with (\ref{cgik})
    \begin{equation}
\psi(x) \to \breve{\psi}(x) =
\exp \left[ \frac{3\, \sigma(x)}{2}  \right] \psi(x)  \,.
\label{phitildew}
\end{equation}
    So differently from scalar field case the Dirac equation for massless case
is conformal invariant without any ambiguities.

\section{Scalar Field in Homogeneous Isotropic Nonstationary Space}
\label{secSPHIS}

    The Lagrangian density for the complex scalar field satisfying~(\ref{mEq})
can be written as
    \begin{equation}
L(x)=\sqrt{|g|}\ [\,g^{ik}\partial_i\varphi^*\partial_k\varphi -(m^2+\xi R)
\varphi^* \varphi \,],
\label{3}
\end{equation}
    where $g={\rm det}\{g_{ik} \}$.
    Varying the action with the Lagrangian~(\ref{3}) with respect to
the variables $ \varphi^*(x)$, $\varphi(x)$ one obtains
the canonical stress-energy tensor
     \begin{equation}
T_{ik}^{\rm \, can}=\partial_i\varphi^* \partial_k\varphi+
\partial_k\varphi^* \partial_i\varphi-g_{ik} |g|^{-1/2}L(x) .
\label{4}
\end{equation}
    Varying the action with the Lagrangian~(\ref{3}) with respect to
the metric $g_{ik}(x)$ one obtains the metric stress-energy tensor
    \begin{equation}
T_{ik}= \frac{2}{\sqrt{ |g|}} \, \frac{\delta S}{\delta g^{ik}} =
T_{ik}^{\rm \, can} - 2 \xi
\left( R_{ik}+\nabla_i\nabla_k-g_{ik}\nabla_j\nabla^j \right)
\varphi^* \varphi \,,
\label{5}
\end{equation}
    where $ R_{ik} $ is Ricci tensor.

    Metric Hamiltonian is
    \begin{equation}
H(\eta)=\int \limits_\Sigma \zeta^i \, T_{ik}(x)\,d \sigma^k,
\label{16}
\end{equation}
    where $ \zeta^i $ is time-like conformal Killing vector,
$ \Sigma $ is the space-time hypersurface.

    In the case of homogeneous isotropic space-time
\begin{equation}
d s^2 = d t^2 - a^2(t)\, d x^2 =
a^2(\eta)\, (d\eta^2 - \gamma_{\alpha \beta} dx^\alpha dx^\beta)
\label{ds}
\end{equation}
    the metric Hamiltonian~(\ref{16}) with $ \zeta^i = (1, 0 , 0, 0) $
has the form
    \begin{equation}
H(\eta)= a^{2}(\eta) \!\! \int \limits_{\eta={\rm const}}\!\!\!T_{00}(x) \,
\sqrt{\gamma}\ d^{3}x,
\ \ \ \ \  \gamma={\rm det}\{\gamma_{\alpha \beta} \},
\label{16w}
\end{equation}
    and the Hamiltonian with canonical stress-energy tensor~(\ref{4})
has the form
    \begin{equation}
H(\eta)=\int \limits_\Sigma \zeta^i \, T_{ik}^{\rm \, can}(x)\,d \sigma^k
= a^{2}(\eta) \!\! \int \limits_{\eta={\rm const}}\!\!\!T_{00}^{\rm \, can}(x) \,
\sqrt{\gamma}\ d^{3}x.
\label{16ww}
\end{equation}
    As it was shown in Ref.~\refcite{Imamura60} for scalar field with minimal
coupling ($\xi = 0$) one obtains infinite result for particle density created
in the expanding Universe (see, also, Ref.~\refcite{Parker69}).

    However as it was shown in Ref.~\refcite{GribNuclPhys69} the method of
the diagonalization of the metrical Hamiltonian of the scalar field with
conformal coupling gives the finite density of created particles.

    It is important to note that for the spinor particles the result for
the number of created particles is always finite.

    But for nonconformal scalar field the using of metric Hamiltonian gives
infinite density of created particles.
    It is noted in Refs.~\refcite{BD,Fulling79}.
    Nonconformal case is important because it is widely used in
inflationary model of the early Universe,\cite{Linde}
massive vector mesons\cite{GMM} (longitudinal components)
and gravitons\cite{GrishchukY80} satisfy equations of this type.
    In case of the scalar field with self interaction in general it is
impossible to conserve conformal invariance not only of the effective action
(the conformal anomaly) but of the action itself.\cite{BD}

    This problem was solved for the nonconformal fields in the Hamiltonian
diagonalization method in Refs.~\refcite{Pavlov00,Pavlov02}.
    One can obtain finite density if one adds to the Lagrangian of nonconformal
field (for example, in case $\xi =0$) some special term in the form of
the divergence of 4-vector
    \begin{equation}
L^{\Delta}(x)=L(x)+ \frac{\partial J^i}{\partial x^i} , \ \ \ \
J^i = \left( \sqrt{\gamma}\, c\,\tilde{\varphi}^* \tilde{\varphi},
\, 0, \, 0, \, 0 \right),
\label{02}
\end{equation}
$$
\tilde{\varphi}(x)= a\,\varphi(x), \ \ \ \ \ \
    c=\frac{a'}{a}. 
$$
    One can construct Hamiltonian as canonical for
the variables $\tilde{\varphi}(x)$.
    Corresponding Hamiltonian density is
    \begin{equation}
h(x)=\tilde{\varphi}'\,
{\displaystyle \frac{\partial L^{\Delta}}{\partial \tilde{\varphi}'}} +
\tilde{\varphi}^{* \prime}\,
{\displaystyle \frac{\partial L^{\Delta}}{\partial \tilde{\varphi}^{* \prime}}}
-L^{\Delta}(x) .
\label{HDens}
\end{equation}
    For the case $\xi =1/6$ it leads to the equality of the canonical Hamiltonian
with metrical one.

\section{Quantization and Particle Creation}
\label{secQPC}

    The scalar field is represented as an integral in complete solutions
of the wave equation in curved space-time
\begin{equation}
\tilde{\varphi}(x)=\int \! d\mu(J) \left[ \tilde{\varphi}{}^{(+)}_J \,
a^{(+)}_J + \tilde{\varphi}{}^{(-)}_J \, a^{(-)}_J \right] ,
\ \ \ \tilde{\varphi}{}^{(+)}_J (x) =\frac{g_J(\eta)}{\sqrt{2}}\,
\Phi^*_{\!J}({\bf x}),
\label{fff}
\end{equation}
where $d\mu(J)$ is a measure on the set of quantum numbers,
    \begin{equation}
 g_\lambda''(\eta) + \Omega^2(\eta) g_\lambda(\eta) =0,
\label{10}
\end{equation}
    \begin{equation}
\Omega^2(\eta)=\left( m^2 -\Delta \xi R\, \right) a^2 + \lambda^2,
\ \ \ \ \ \ \Delta \xi = \xi_c - \xi,
\label{010}
\end{equation}
    $ \Phi_{J}({\bf x})$ are the eigenfunctions of the Laplace-Beltrami operator
on 3-space with metric $\gamma_{\alpha \beta} $.

    Fock operators $\stackrel{*}{a}\!{\!}^{(-)}_{\bar{J}}$ and  $a^{(-)}_J$
satisfy standard commutation relations.
    Then the Hamiltonian corresponding to~(\ref{HDens}) is
\begin{equation}
H(\eta)=\int \! d\mu(J) \! \left[ E_J(\eta) \!
\left(\stackrel{*}{a}\!{\!}^{(+)}_J a^{(-)}_J +
\stackrel{*}{a}\!{\!}^{(-)}_{\bar{J}} a^{(+)}_{\bar{J}} \right) \! +
F_J(\eta) \stackrel{*}{a}\!{\!}^{(+)}_J a^{(+)}_{\bar{J}}\! +
F^*_J(\eta) \stackrel{*}{a}\!{\!}^{(-)}_{\bar{J}} a^{(-)}_J  \right]\!,
\label{Hhh}
\end{equation}
    \begin{equation}
E_J(\eta)=\frac{1}{2} \left[\, |g_J'|^2+ \Omega^2 |g_J|^{\mathstrut 2} \,\right]
\ , \ \ \ F_J(\eta)= \frac{\vartheta_{\!J}}{2} \left[ g_J'{}^{\! 2} +
\Omega^2  g_J^{\mathstrut 2}  \right]  ,
\ \ \vartheta_{\!J} \!=\! \pm 1.
\label{EJFJ}
\end{equation}
    The initial conditions  corresponding to
the diagonal form at $\eta_0 $ are
    \begin{equation}
g_\lambda'(\eta_0)=i\, \Omega(\eta_0)\, g_\lambda(\eta_0) , \ \ \
|g_\lambda(\eta_0)|= 1/\sqrt{\Omega(\eta_0) } .
\label{40}
\end{equation}
    Hamiltonian diagonalization is realized by Bogoliubov transformations
    \begin{equation}
\left\{  \begin{array}{c}
a_J^{(-)}=\alpha^*_J(\eta) \,b^{(-)}_J(\eta)-
\beta_J(\eta)\vartheta_J\, b^{(+)}_{\bar{J}}(\eta) \,,  \\[3mm]
\stackrel{*}{a}\!{\!}_J^{(-)}=\alpha^*_J(\eta) \,
\stackrel{*}{b}\!{\!}^{(-)}_J\!(\eta)-
\beta_J(\eta)\vartheta_J \,\stackrel{*}{b}\!{\!}^{(+)}_{\bar{J}}\!(\eta) \,,
\end{array} \right.
\label{26}
\end{equation}
    where
$|\alpha_J(\eta_0)|=1, \ \ \beta_J(\eta_0)=0 $,\ \ \
$ |\alpha_J(\eta)|^2-|\beta_J(\eta)|^2=1   $.

   The density  of the created pairs of particles  
corresponding to diagonal form of Hamiltonian is
    \begin{equation}
n(\eta)=\frac{1}{2 \pi^2 a^3(\eta)}
\int d\lambda\,\lambda^2 \,S_\lambda(\eta),
\ \ \ \
S_\lambda(\eta)= \frac{|g^\prime_\lambda|^2+\Omega^2 |g_\lambda|^2}{4\Omega}
- \frac{1}{2} .
\label{33}
\end{equation}
    For the conformal and nonconformal cases with arbitrary $\xi$
one has\cite{GMM80,Pavlov00}
$ S_\lambda(\eta) \sim \lambda^{-6} ,\ \lambda \to \infty $
and the density of created particles is finite.

\subsection{Scalar particle creation (numerical results)}
\label{secSPCN}

\begin{figure}[h]
\centering
   \includegraphics[width=0.45\textwidth]{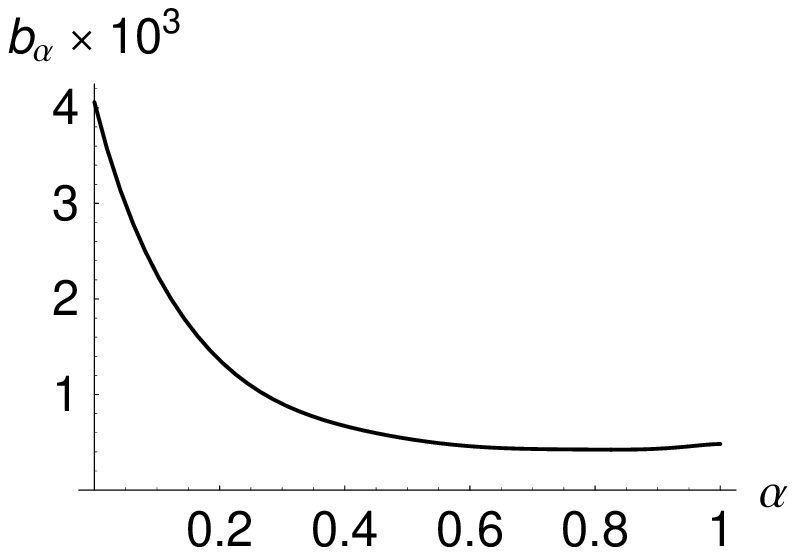} \ \
\includegraphics[width=0.45\textwidth]{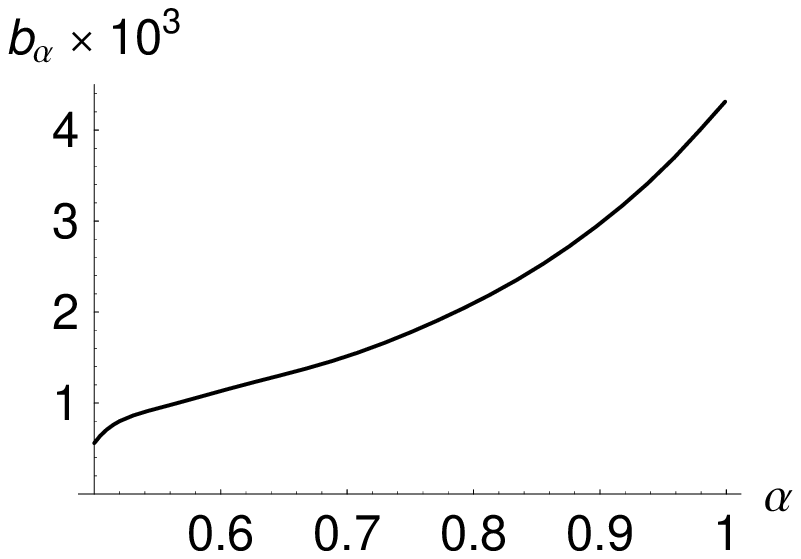}
\caption{$b_\alpha$ for scalar particles with the
conformal (at the left) and minimal (at the right) coupling to curvature.}
\label{grBq}
\end{figure}

    For conformal particle creation with nonzero mass and the scale factor $a$
of the Friedman Universe one obtains the result for the dependence of
the number of particles created in the Lagrange volume on
the parameter~$\alpha$ as
    \begin{equation}\label{Na}
a(t)=a_0 t^\alpha, \ \ \ \ \
N(t)= \left( \frac{a(t_C)}{t_C} \right)^3 b_{\alpha} ,
\end{equation}
    where $t_C= 1/m$ is Compton time.
    On Fig.~\ref{grBq} (at the left) the dependence of $b_\alpha$ on $\alpha$
for conformal scalar particles is shown.

    For nonconformal scalar particle creation one can also receive finite
result for special conditions.
    The choice of initial time is made as $t_0$:
$$
\Delta \xi =0 \ \ \Rightarrow \ \ t_0=0,
$$
$$
\Delta \xi R > 0 \ \ \Rightarrow \ \ t_0 \ \ \ \mbox{from} \ \ \
m^2-\Delta \xi R(t_0)=0.
$$
    For example, in the case $a(t)= a_0 t^\alpha$ with $\alpha > 1/2$
one has $t_0 = t_C \sqrt{\Delta \xi \, 6 \, \alpha (2 \alpha -1 )}$.
    On Fig.~\ref{grBq} (at the right) the dependence of $b_\alpha$ on $\alpha$
for scalar particles with mass and minimal coupling is shown.

    One can ask what is the reason for the ansatz~(\ref{02}) leading
to finite results?
    The reason is that it leads to the consequence that our Hamiltonian is
connected with the canonical Hamiltonian in conformally connected static
case for the conformally transformed field with varying mass depending on time.

\section{Superheavy Particles in the Early Universe and Origination of
Dark and Visible Matter}
\label{secSHPU}

    The gravity of Friedmann radiation dominated Universe $a(t)=a_0 \sqrt{t}$
creates
    \begin{equation}
N=n^{(s)}(t)\,a^3(t)=b^{(s)}\,M^{3/2}\,a_0^3
\label{NbMw}
\end{equation}
    pairs of particles inside the Lagrange volume, where
$b^{(0)} \approx 5.3 \cdot 10^{-4}$ for scalar and
$b^{(1/2)} \approx 3.9 \cdot 10^{-3}$ for spinor particles.\cite{GMM}

    The physical meaning of the effect of particle creation in the early
Universe was underestimated:
in Refs.~\refcite{Zeldovich75,Weinberg77} one can read that the effect is
physically negligible,
and Ref.~\refcite{BD} does not mention the numerical results at all.
    All this occurred due to numerical estimates made only for particles of
low mass observed today in our laboratories.

    However if one takes the mass of the Grand Unification order
(or the mass of the inflaton) the number of created particles occurs
to be close to the observable Eddington-Dirac value.
    By calling these particles $X$-particles and supposing them to decay
in visible particles in the time not far from the time of their
creation one can obtain the observable particles number.\cite{GMM,Grib95}

    Superheavy particles are nonstable at the time of Grand Unification
symmetry but become stable when the symmetry is broken.
    For some time $ t_X $ there is an era of going from the
radiation dominated model to the dust model of superheavy particles:
    \begin{equation}
t_X\approx \left(\frac{3}{64 \pi \, b^{(s)}}\right)^2
\left(\frac{M_{Pl}}{M}\right)^4 \frac{1}{M}  .
\end{equation}
    If $M \sim 10^{14} $\,Gev,
$ t_X \sim 10^{-15} $\,s for scalar
and $\ t_X \sim 10^{-17} $\,s for spinor particles.

    It is possible that some of these $X_L$ survive up to modern time
and exist as part of cold dark matter.
    Then it is possible to identify the decay of $X_L$-particles in
modern epoch as the UHE cosmic ray events.
    Let us define $d$ --- the permitted part of long living $X$-particles
from condition $ d \cdot \varepsilon_X(t_{rec}) = \varepsilon_{crit}(t_{rec})$,
then
    \begin{equation}
d=\frac{3}{64 \pi \, b^{(s)}}\left(\frac{M_{Pl}}{M_X}\right)^2\,
\frac{1}{\sqrt{M_X\,t_{rec}}} .
\label{d}
\end{equation}
    For $M_X=10^{13} - 10^{14} $\,Gev one has $d \approx 10^{-12} -
10^{-14} $ for scalar,
$d \approx 10^{-13} - 10^{-15} $ for spinor particles.

    Our hypothesis\cite{GribPV08} is
---  the decay of superheavy particles of a dark matter,
which are stable in usual conditions, can occur
in ergosphere of a supermassive, quickly rotating black holes in
active galactic nuclei.

    Some numerical estimates made in Refs.~\refcite{GribPV09,GribPV09b}
show us, that  decays of dark matter superheavy particles in the vicinity of
supermassive black holes of the active galactic nuclei can provide
observed intensity of the ultrahigh energy cosmic rays.

\section{Conclusions}
\label{secConcl}

\hspace*{\parindent}
{\bf 1.}
    There is a widespread opinion due to some authoritative text books
(Ya.\,Zel'dovich -- I.\,Novikov, S.\,Weinberg)
that particle creation in the expanding universe plays negligible role.
This is correct for visible known particles.
    However this occurs to be nontrue for superheavy particles of the mass of
the Grand Unification scale.
    It can be that only these particles were created from gravitation while
all visible today particles appeared due to decay of these primordial particles.

{\bf 2.}
    The idea of Big Bang as some explosion of matter  looks not exact.
It can be that close to singularity as the edge of space --- there was no matter at all
(as it occurs for example in empty anisotropic solution of Einstein equations)
and all matter appeared later.
Singularity is a property of geometry but not of matter.

{\bf 3.}
    Processes of the decay of superheavy particles as particles of dark matter
on ordinary particles can occur in the vicinity of the horizon of
rotating black holes.

\section*{Acknowledgments}

This research is supported by the Russian Foundation for Basic research
(Grant No. 18-02-00461 a).
The work of Yu.V.P. was supported by the Russian Government Program of
Competitive Growth of Kazan Federal University.


\end{document}